\documentclass{ametsocV6.1}

\title{Prithvi-Precip: Integrating Satellite Observations into an Atmospheric AI
Foundation Model for Precipitation Forecasting}

\usepackage{booktabs}

\authors{
  Simon Pfreundschuh\aff{a}\correspondingauthor{Simon Pfreundschuh, simon.pfreundschuh@colostate.edu},
  Christian D. Kummerow\aff{a},
  Johannes Schmude\aff{b},
  Sujit Roy\aff{c},
  Rahul Ramachandran\aff{d},
  Tsengdar Lee\aff{e},
  Valentine Anantharaj\aff{f},
  Katherine H. Breen\aff{g, h}
}

\affiliation{
  \aff{a}{Colorado State University, Fort Collins, CO, USA}\\
  \aff{b}{IBM Research} \\
  \aff{c}{Earth System Science Center, The University of Alabama in Huntsville, Huntsville,  AL, USA; }
  \aff{d}{NASA Marshall Space Flight Center, Huntsville, AL, USA}\\
  \aff{e}{NASA Headquarters, Washington, DC, USA}\\
  \aff{f}{National Center for Computational Sciences / Oak
Ridge National Laboratory, Oak Ridge, TN, USA;}\\
  \aff{g}{Morgan State Univeresity, Baltimore, MD, USA}\\
  \aff{h}{Global Modeling and Assimilation Office, NASA Goddard Space Flight Center, Greenbelt, Maryland}\\
}

\abstract{%
Accurate precipitation forecasting remains one of the most challenging problems
in weather prediction. While recent AI weather prediction (AIWP) systems have
achieved substantial improvements in medium-range forecasting skill,
precipitation often remains a secondary target and is commonly learned from
reanalysis datasets that contain considerable uncertainty. In this work, we
investigate two complementary strategies for improving AI-based precipitation
forecasts. Building on the Prithvi-WxC foundation model, we develop
Prithvi-Precip, a global precipitation forecasting system, and examine (1) the
impact of training targets derived from satellite-based precipitation estimates
rather than reanalysis fields and (2) the direct assimilation of satellite
observations into the forecasting model.\\%
We systematically evaluate key design choices for finetuning the Prithvi-WxC AI
foundation model for precipitation forecasting. We find that autoregressive
rollout training produces substantially more accurate forecasts than direct
conditioning on forecast lead time. Using independent radar-based precipitation
estimates for evaluation, we show that training on satellite-derived
precipitation targets yields improved forecast accuracy relative to training on
MERRA-2 precipitation fields. Furthermore, direct ingestion of satellite
observations provides additional improvements at short lead times, with the
largest gains occurring in tropical and subtropical regions.\\%
Together, these advances enable Prithvi-Precip to substantially improve upon
directly comparable precipitation forecasts from the Goddard Earth Observing
System. Our results highlight the potential of improved precipitation targets
and the direct integration of satellite observations as promising pathways for
advancing medium-range AI precipitation forecasting.
}

\begin{document}

\maketitle

%
%
%
\statement

Precipitation remains one of the most difficult weather variables to forecast
accurately, even with recent advances in artificial intelligence. This study
shows that AI precipitation forecasts can be improved by using satellite-derived
precipitation estimates as training targets and by directly incorporating
satellite observations into the forecasting model. The study thus identifies
practical strategies for improving global AI precipitation prediction and
demonstrate that both the quality of the training data and the use of current
satellite observations are potential avenues for advancing medium-range
precipitation forecasting.

%
\section{Introduction}
\label{sec:intro}

Precipitation is central to human life and activity: it sustains freshwater
resources, underpins agriculture and energy systems, and drives extreme events
that can threaten lives and economies. Yet, producing accurate quantitative
precipitation forecasts using numerical weather prediction (NWP) models remains
challenging owing to the complex microphysical processes involved in its
formation \citep{morrison_2020}, high spatiotemporal variability, and poor
representation in the initial states of NWP models
\citep{li_convolutional_2022}.

AI weather prediction (AIWP) models learn to forecast weather directly from
atmospheric (re-)analyses allowing them to learn the evolution of the atmosphere
directly from data instead of the explicit physical modeling applied in
conventional NWP models. AI-based weather forecasting has made important
advances in the last year and a range of models now exceeds conventional
baselines by a significant margin \citep{lang_aifs_2024}.

However, precipitation has so far not been a primary focus of many recent global
AI weather prediction (AIWP) efforts. Early models, such as Pangu-Weather
\citep{bi_pangu-weather_2022} and Stormer \citep{nguyen_scaling_2024}, did not
predict precipitation explicitly. More recent models, including FourCastNet
\citep{pathak_fourcastnet_2022}, GenCast \citep{price_gencast_2024}, FuXi
\citep{zhong_fuxi-20_2024}, and the AIFS \citep{lang_aifs_2024}, do provide
precipitation forecasts. However, the evaluation of AIWP models still primarily
focuses on large-scale dynamical variables such as temperature and geopotential
height. This is because precipitation remains difficult to evaluate consistently
at global scales because reanalysis products provide only indirect and uncertain
estimates of precipitation \citep{rasp2024weatherbench, radford_2026}.

At shorter lead times, a complementary class of AI-based nowcasting systems has
emerged that focuses specifically on precipitation forecasting using radar- or
satellite-derived precipitation estimates as training targets
\citep{agrawal_operational_2025}. These systems avoid some of the limitations
associated with reanalysis-based precipitation references, but are generally
limited to lead times of less than 12 hours.

Recent work has begun exploring AI forecasting systems that operate more
directly on observations in order to obviate the need for complex data
assimilation systems or overcoming their shortcomings \citep{allen2025end}.
Examples include AIWP forecasts trained to predict reanalysis states directly
from observations \citep{allen2025end} as well as systems operating completely
in observations pace \citep{mcnally2024data, gong2025}.

This work explores approaches to improving medium-range AI precipitation
forecasting. Building on the Prithvi-WxC foundation model
\citep{schmude_prithvi_2024}, we finetune a global forecasting system to predict
precipitation at six-hour intervals up to 96 hours into the future.
Specifically, we investigate two strategies for improving precipitation
forecasts: (1) improving the accuracy of the target precipitation estimates and
(2) directly integrating satellite observations into the forecasting model.

To explore the impact of the accuracy of the target precipitation estimates, we
train the forecasts using precipitation estimates from the Modern-Era
Retrospective Analysis for Research and Applications
\citep[MERRA2]{gelaro_modern-era_2017} and satellite-based precipitation
estimates from the Integrated-Multi-Satellite Retrievals for the Global
Precipitation Measurement \citep[IMERG]{huffman_integrated_2020} V07 and evaluate
estimates against independent precipitation estimates gauge-corrected
ground-based radars over CONUS and direct gauge measurements over Brazil.

For the integration of satellite observations, we propose a sensor agnostic
encoding mechanism that ingests raw satellite observations directly into the
forecast without requiring sensor-specific encoding. This allows training on
satellite observations from a wide range of sensor types including passive
microwave sensors in low-earth orbit and visible and infrared observations
from geostationary platforms.

Through validation against independent gauge-corrected radar and gauge
observations, we show that both interventions improve precipitation forecast
skill relative to forecasts trained on MERRA-2 reanalysis precipitation alone
and on baseline forecasts from the Goddardd Earth Observing System
\citep[GEOS]{suarez2005documentation} forward processing (FP) weather
forecasting system. Improvements from the improved precipitation data reference
are substantial across the full assessed 96-hour forecast range, while the
impact from satellite observations is strongest at short lead times but
diminishes at longer lead times.

\section{Data and Methods}

\label{sec:methods}

\subsection{The Prithvi-WxC AI Foundation Model}

\begin{figure}[h]
 \centerline{\includegraphics[width=37pc]{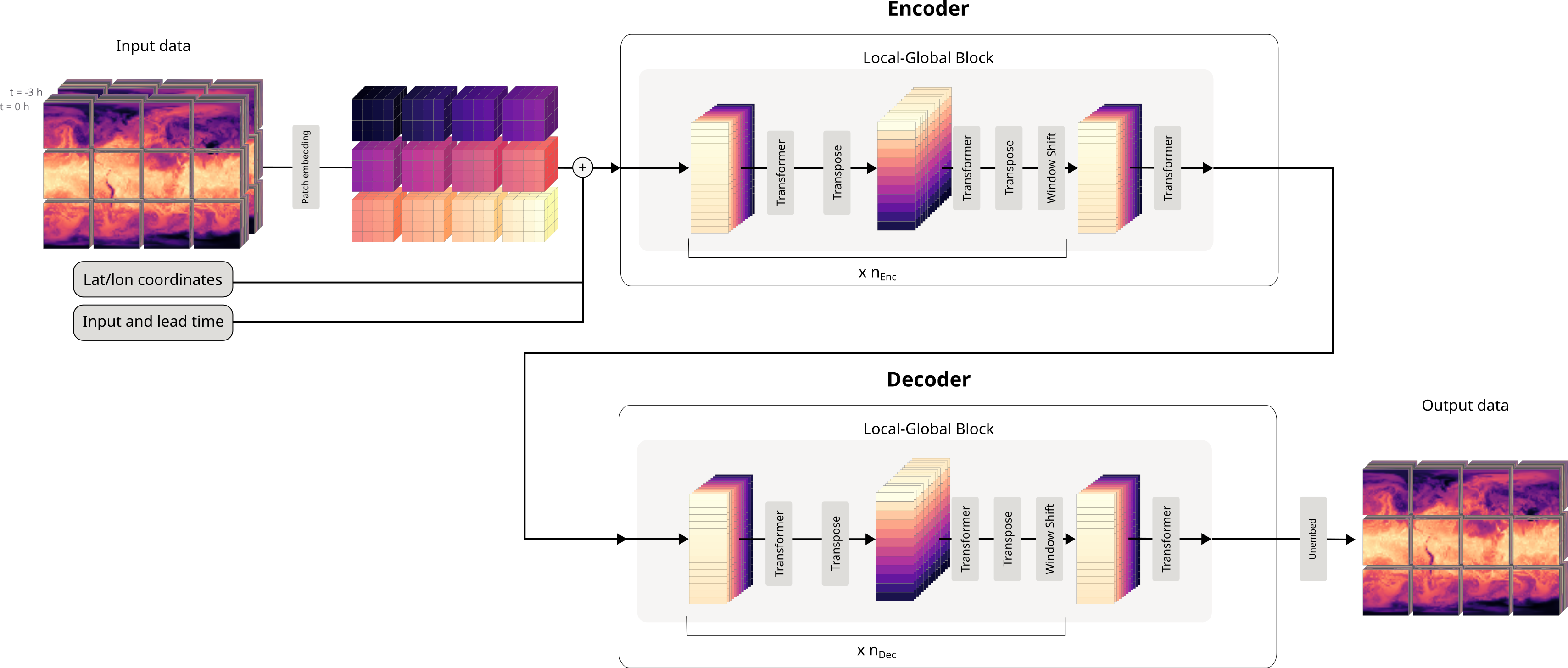}}
 \caption{
Architecture of the Prithvi-WxC AI Foundation Model. The model consists of a
transformer-based encoder and decoder. Both components are built from a stack of
local-global blocks that alternately operate on the latent state within
individual spatial tiles and across tiles, enabling information exchange across both
local and global scales.
 }\label{fig:prithvi_wxc}
\end{figure}

The Prithvi-WxC model \citep{schmude_prithvi_2024} is a transformer-based global
geophysical AI foundation model. The model has been pre-trained on the MERRA-2
reanalysis dataset to predict present and future states from masked input data.
The model architecture, which is illustrated in Fig.~\ref{fig:prithvi_wxc}, is
inspired by the Hiera \citep{ryali_hiera_2023} and MaxViT \citep{tu2022_maxvit}
models. The primary input data consist of two time steps of MERRA-2 data
comprising 20 single-level and 10 vertically-resolved variables at 14 vertical
levels. Auxiliary input data includes the targeted lead time, the climatology
corresponding to the targeted forecast date, the spatial coordinates of the data
grid, and the time difference between the two input time steps. The output of
the model are the anomalies of the 160 dynamic atmospheric fields for the
targeted lead time with respect to the climatology.

This work uses two versions of the Prithvi-WxC model. The large 2.7 Billion
parameter model described in \citet{schmude_prithvi_2024} and a small version
with only 280 million parameters. This work uses the large Prithvi-WxC model
that went through additional rollout training whereas the small model used here
has not been trained tuned for rollout.

The Prithvi-WxC pre-training uses the same regular latitude–longitude grid as
the the MERRA-2 reanalysis dataset, which has a resolution of 0.67$^\circ$ in
the zonal dimension and 0.5$^\circ$ in the meridional direction. Internally, the
Prithvi-WxC model splits the input data into horizontal tiles along the zonal
and meridional dimensions. The model alternatingly applies transformer layers to
all grid points within a given tile and globally across the corresponding grid
points in each tile to allow for both local and global information exchange. To
avoid tiling artifacts, the model additionally applies shifting in between
consecutive tile-local operations similar to the Swin architecture
\citep{Liu_2021_ICCV}.

\subsection{Forecasting Precipitation with the Prithvi-WxC model}

Since precipitation is not an output variable of the original Prithvi-WxC model,
we attach an additional head consisting of a pixel-wise four-layer MLP with GELU
activation functions and layer normalization to the output of the Prithvi-WxC
model. The MLP is applied to the raw model output prior to the scaling by
layer-wise standard deviations and the addition of the climatological means
\citep{schmude_prithvi_2024}. This extended version of the Prithvi-WxC model
thus produces two outputs at the requested lead time, the 160-variable
atmospheric state and a 2D global precipitation field. The full atmospheric
state is retained to enable autoregressive predictions using rollout. The
precipitation field is not fed back into the model during rollout.

\subsection{Integrating Satellite Observations}

A central challenge for integrating satellite observations into an AIWP model is
the heterogeneity of the observations. The record of meteorological satellite
observations is derived from a diverse set of sensors that have evolved over
decades of satellite remote sensing, each providing measurements with different
channel counts, spectral characteristics, spatial resolutions and sampling. This
diversity poses a fundamental difficulty for conventional machine-learning
approaches, which typically assume a fixed set of input features for model
inputs. In addition, satellite observations are spatially sparse with
observations from a single sensor during a typical forecast timestep covering
only a fraction of the globe.

\begin{figure}[h]
 \centerline{\includegraphics[width=37pc]{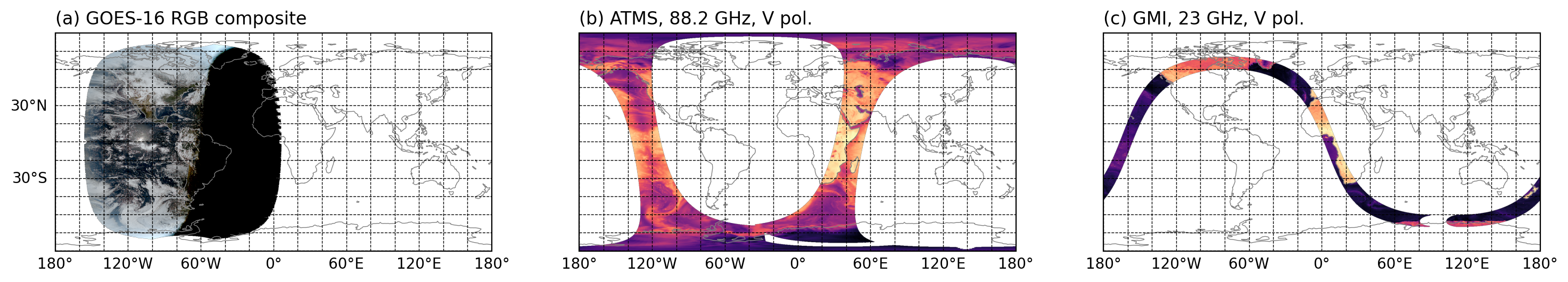}}
 \caption{
Examples of meteorological satellite observations from three different sensors.
Panel (a) shows an RGB composite from the Advanced Baseline Imager
\citep{schmit2005introducing} on the GOES-16 geostationary satellite. Panel (b)
shows quasi-vertically polarized observations from the 88.2 GHz channels of the
Advanced Technology Microwave Sensor (ATMS, \citealp{goldberg_advanced_2006}).
Panel (c) shows vertically polarized observations from the 23-GHz channel on
the GPM Microwave Imager (GMI, \citealp{draper_global_2015}). Black dashed lines
show the tiling applied by the Prithvi-WxC model.
}\label{fig:satellite_observations}
\end{figure}

Figure~\ref{fig:satellite_observations} illustrates the challenges of
integrating heterogeneous satellite observations into AIWP models using
representative examples from three classes of meteorological satellite sensors.
Geostationary satellites (Panel (a)) remain at a fixed position relative to the
surface of the Earth, which allows them to continuously observe much of the
hemisphere below them. While geostationary observations can provide continuous
observations of a significant portion of the globe, their observations are
limited to visible and infrared wavelengths, which due to their sensitivity to
small cloud hydrometeors are unable to penetrate deeply into clouds and thus
 provide only limited information on precipitation close to the ground.

Polar-orbiting satellites (Panel (b)) sample the entire globe through
successive, shifting swaths with limited instantaneous coverage. Their low
altitude makes these sensors suitable for microwave sensors, which are less
sensitive to small cloud particles and can provide a more direct signal
from surface precipitation.

Finally, specialized science instruments —such as the GPM Microwave Imager
\citep{draper_global_2015} aboard the Global Precipitation Measurement (GPM,
Panel (c)) — offer still narrower swaths but deliver higher-resolution and more
information-rich measurements than most operational meteorological sensors.

In order to integrate satellite observations into the Prithvi-WxC model in a
generalized fashion, each sensor channel is handled as a separate observation
layer. Furthermore, to handle the sparsity of the observations on the
latitude--longitude grid employed by the Prithvi-WxC model, observations
are stored as tiles matching the internal tiling of the Prithvi-WxC model and
empty tiles are discarded. Using this unified approach allows us to build an extensive
observation record across sensor generations.

\subsubsection{Observation Encoder}

\begin{figure}[h]
 \centerline{\includegraphics[width=37pc]{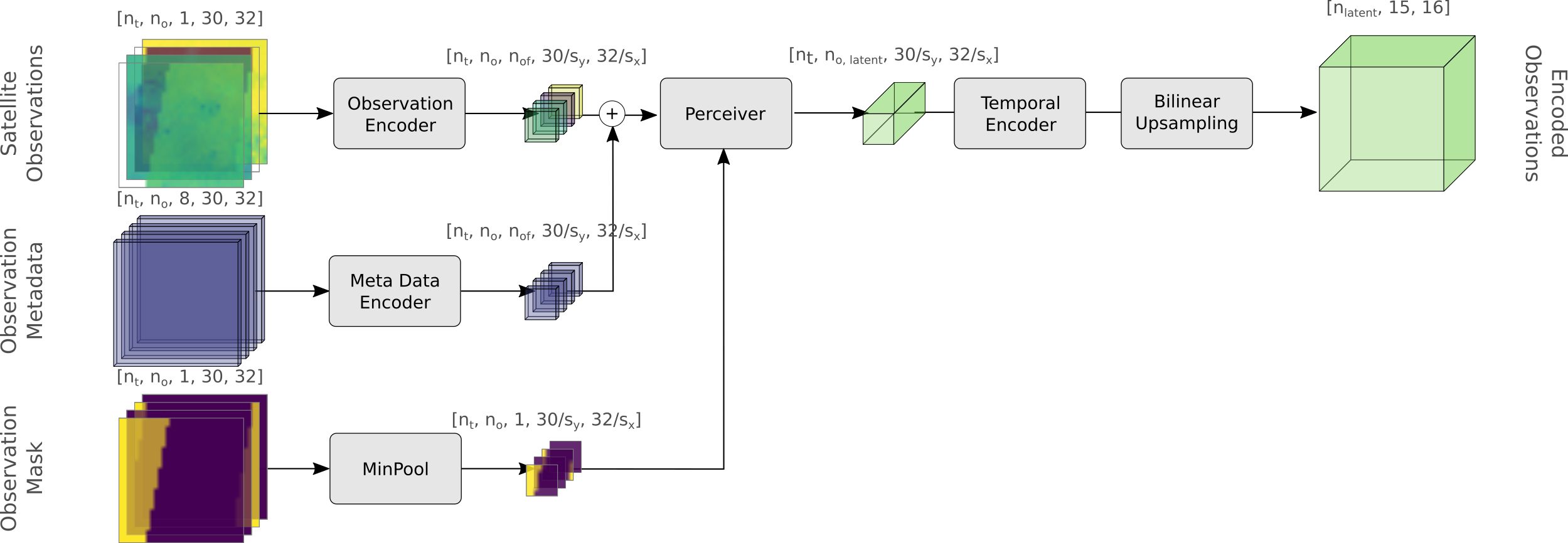}}
 \caption{
Architecture of the observation encoder used by the original Prithvi-Precip model. Each
observation channel is stored as an independent observation layer with
corresponding metadata and observation mask. Observations and metadata are
encoded on the tile level to observation tokens of size $s_y \times s_x$ with
$s_y=6, s_x=4$ and then mapped from an arbitrary-length sequence of observations
layers to a fixed-dimensional latent space with
using cross attention. The encoded observations from the two input timesteps are
then time-encoded and upsampled to the original resolution.
 }\label{fig:observation_encoder}
\end{figure}

To incorporate the observations into the Prithvi-WxC model, an observation
encoder is added to the model. The observation encoder -- displayed in
Fig.~\ref{fig:observation_encoder} -- encodes observations at each tile
independently, mapping the variable-length sequence of observation layers
available at each tile to a fixed-size latent space. To allow the observation
encoder to distinguish between different observation layers, the wavelength,
polarization, and observation time relative to a three-hour input window are
provided as auxiliary input to each observation layer. The observation and
metadata sequences for each tile are encoded into observation tokens with a
patch size of 6 x 4 pixels. The large patch size was chosen to reduce the memory
footprint required by the observation encoder while the asymmetry is due to the
tile size used by the Prithvi-WxC model which prohibits downsampling by a factor
of 4 along both height and width. The arbitrary-length sequences of observation
layers are combined with the corresponding metadata and encoded into a
fixed-dimensional latent space using cross-attention with a single learnable query
similar to the approach used in the Perceiver architecture \citep{jaegle2022}.
The encoded observations from the two input timesteps are then combined using a
temporal encoder and upscaled to match the token size used internally by the
Prithvi-WxC model.

\begin{figure}[h]
 \centerline{\includegraphics[width=37pc]{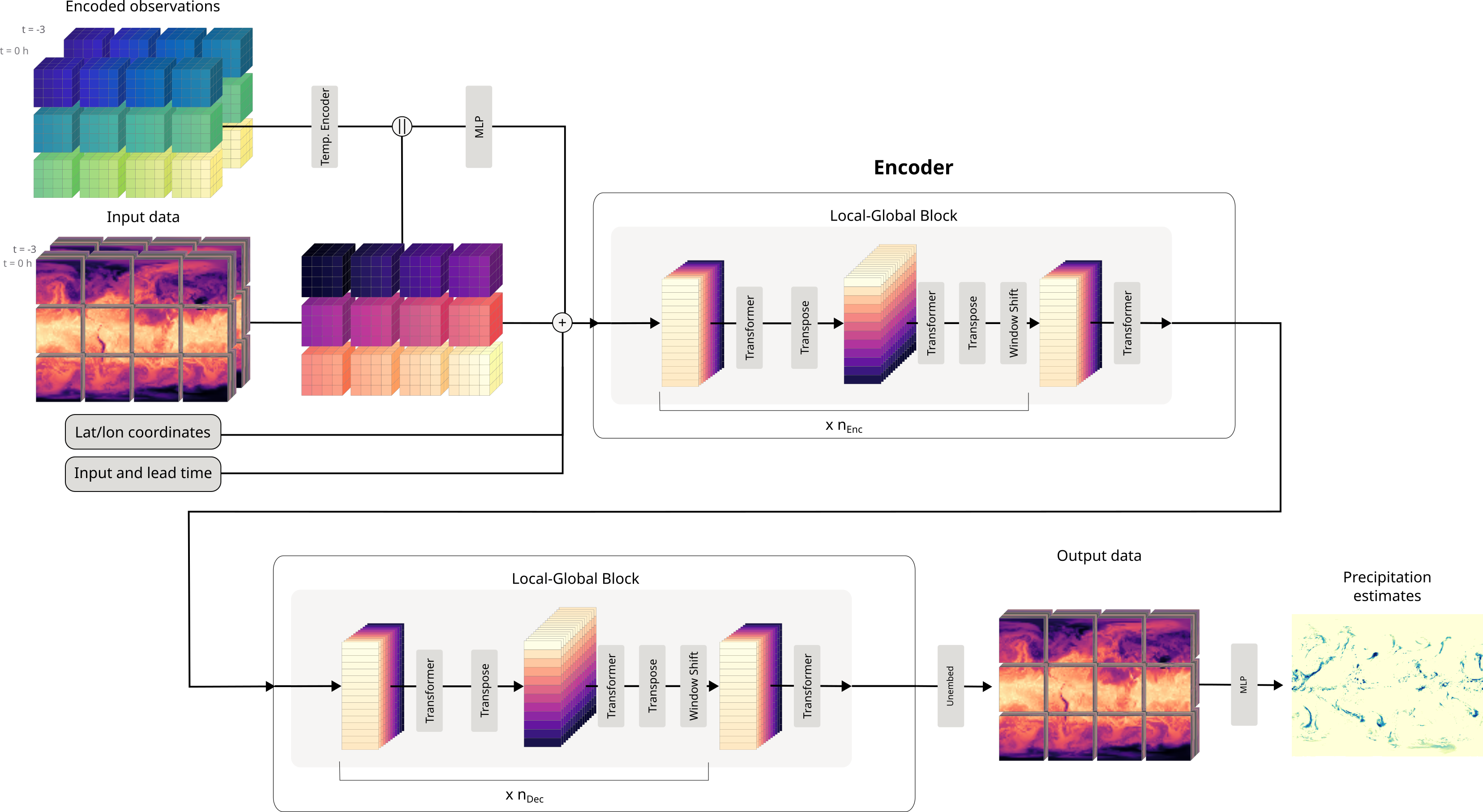}}
 \caption{
Architecture of the final Prithvi-WxC Precip model to forecast precipitation
using satellite observations. Observations encoded using the observation encoder
shown in Fig.~\ref{fig:observation_encoder} are merged with the encoded model by
calculating a residual correction using a two-layer MLP applied token-wise to
the concatenated latent model state and observations. An additional 4-layer MLP
is used to produce precipitation estimates from the unembedded output of the
Prithvi-WxC backbone.
 }\label{fig:prithvi_precip}
\end{figure}

The encoded observations are then merged with the encoded atmospheric state of
the Prithvi-WxC model. This is done using a two-layer MLP that takes the encoded
atmospheric state concatenated with the encoded observations and the current
unrolling step and calculates a residual correction that is added to the encoded
atmospheric state. The resulting architecture of the Prithvi-WxC Precip model is
shown in \ref{fig:prithvi_precip}.

\subsection{Training}

The Prithvi-Precip model is trained using the same MERRA-2 input data used for
the pretraining of the Prithvi-WxC model. Precipitation estimates from MERRA-2
and and IMERG V07 are downsampled to the spatial resolution of the MERRA-2 grid
and a temporal resolution of 3 h are used as training targets. The model is
trained to predict 32 quantiles of the predictive distribution of precipitation
at each grid point using a quantile loss function
\citep{pfreundschuh_neural_2018}.

To determine the best strategy for performing finetuning the Prithvi-WxC model
for precipitation forecasting, we evaluate two different forecasting strategies:
autoregressive rollout and continuous forecasting \citep{nguyen_scaling_2024}.
For the autoregressive fine-tuning, the model is rolled out to predict
precipitation at six six-hour steps up to 36 h into the future. Model parameters
are optimized using the AdamW \citep{loshchilov_decoupled_2019} optimizer using
only the quantile loss of the predicted precipitation as training objective.
Training uses data from the years 2000–2020 with an epoch consisting of 10\% of
all available input timesteps from the training period. The model is trained for
100 epochs during which the learning rate is linearly warmed up over the first 5
epochs to $5 \cdot 10^{-4}$, after which it follows a cosine-annealing schedule
with restarts after 13, 26, and 52 epochs.

For continuous forecasting, the model is trained to directly predict
precipitation up to 96 hours ahead with the desired lead time provided as part
of the model input. Since continuous forecasting is less demanding
computationally, the dataset is only subsampled by a factor of 2 per epoch.
During initial experiments we have observed lead-time dependent overfitting with
forecasts of lead times exceeding 48 h starting to overfit while the accuracy of
forecasts for shorter lead times are still improving. We therefore apply
aggressive data augmentation for training the continuous forecasts. More
specifically, we apply random zonal rolling, meridional flips and scaling of the
input data. For meridional flips we also adapt the sign of the meridional wind
fields accordingly. For the random flips and rolls, we keep the longitude and
latitude fields unchanged since the Prithvi-WxC is invariant to these
transformations when they match the tile size.

To train the final Prithvi-WxC Precip model with satellite observations, the
model is first fine-tuned using only reanalysis inputs following the procedures
described above. The observation encoder is then added and training is repeated,
initializing all shared weights from the reanalysis-only model. The
encoded reanalysis and observation inputs are each randomly dropped with a
probability of 20\%, ensuring that the resulting model can operate with either
input source alone. Additionally, a layer-wise dropout with a 20\% probability
is applied to the observation layers to further increase robustness to missing
satellite observations. For each forecast, 48 observation layers are randomly
chosen from the observations available for each 3-hours window during the input
data window leading to a total of 192 observations layers across two time steps.

\subsection{Training and Validation Data}

The training of the Prithvi-Precip model uses the same reanalysis-based
atmospheric input data as the underlying Prithvi-WxC model. To provide
additional, direct observational constraints on the initial state of the
atmosphere, the reanalysis based input data is complemented with raw satellite
observations from passive microwave and geostationary sensors.

\subsubsection{Precipitation Reference Data}

To assess the impact of the precipitation target data on the resulting forecast
model, we compare precipitation estimates from two different datasets. Since the
Pritvh-WxC model is trained using MERRA-2 data, precipitation estimates from
MERRA-2 constitute a natural choice as reference data for precipitation
forecasts. However, since reanalysis precipitation is not directly constrained
by observations and often struggles to accurately represent precipitation
\citep{lavers2022evaluation}, we use precipitation estimates from Version 07 of
the Integrated Multi-Satellite Retrievals for GPM (IMERG) dataset
\citep{huffman_integrated_2020}. IMERG combines precipitation estimates from
multiple satellite sensors to produce global precipitation estimates at
30-minute temporal resolution. The IMERG Final run used here additionally
applies a monthly gauge corrections over land surfaces.


\subsubsection{Independent Validation Data}

To independently evaluate forecast skill, we use gauge-corrected precipitation
estimates from the NOAA Multi-Radar Multi-Sensor (MRMS) dataset
\citep{smith_multi-radar_2016}. MRMS combines observations from the NEXRAD
weather radar network with surface gauge measurements to provide high-resolution
precipitation estimates over CONUS. Gauge-corrected radar precipitation
estimates are among the most reliable spatially continuous precipitation
estimates and are widely used for the evaluation of satellite precipitation
retrievals and precipitation forecasts.

To complement the ground-based radar data over CONUS, we use direct gauge
measurements from the Brazilian Instituto Nacional de Meteorologia (INMET)
\citep{inmet_data}. The data comprises measurements from 524 automated stations
providing precipitation measurements every hour.

\subsubsection{Baseline Forecasts}

We use precipitation forecasts from the GEOS-FP forecasting system from the
period March–August 2025 as a conventional baseline. Because GEOS-FP employs the
same forecast model and 3D-Var data assimilation system used to generate the
MERRA-2 reanalysis, it provides the most direct comparison for the AI forecasts
produced by the Prithvi-Precip model.

As strong AIWP baseline, we include experimental operational forecasts from the
AIFS \citep{lang_aifs_2024}. These forecasts are initialized from the analysis
produced by the IFS data assimilation system, one of the leading global
numerical weather prediction systems. Consequently, AIFS constitutes a
particularly strong benchmark against which to evaluate the Prithvi-Precip
forecasts.

\subsubsection{Satellite Observations}

We use passive microwave observations from multiple sensors of the GPM
constellation beginning in the year 2000, including AMSU-B, MHS, ATMS, SSM/I, SSMIS, GMI.
These observations are extracted from the Level-1C inter-calibrated brightness
temperature products produced by the GPM mission. The Level-1C products retain
channels primarily sensitive to precipitation-related hydrometeors and water
vapor while excluding traditional temperature-sounding channels available in the
original sensor observations.
Because passive microwave observations provide only intermittent spatial
coverage, we additionally use observations from geostationary satellites. For
global infrared observations, we use gridded window-channel brightness
temperatures from the CPC Gridded IR dataset \citep{noaa_cpc_ir}. We further
incorporate visible, water vapor, and infrared observations from the GridSat-B1
\citep{knapp2011_gridsat} and GridSat-GOES datasets
\citep{knapp2018_gridsat_goes}. For inference in recent years not covered by the
GridSat-GOES dataset, equivalent observations are extracted directly from the
current GOES archive \citep{noaa_goes_open_data}.
A list of all observation channels used by the Prithvi-Precip model is provided
in Table~\ref{tab:sensors}.

\begin{table*}[hbpt]
\centering
\caption{Overview of the satellite observations used to drive the Prithvi-Prithvi forecasts and their sources}
\label{tab:sensors}
\small
\begin{tabular}{p{2.2cm} p{5.0cm} p{2.2cm} p{3.0cm}}
\hline
\textbf{Sensor/Dataset} & \textbf{Channel Frequencies/Wavelengths} & \textbf{Temporal Coverage} & \textbf{Source} \\
\hline
SSMI &
19.35 GHz, 22.23 GHz, 37 GHz, 89.0 GHz &
2000--2015 &
\citet{gpm_l1c_f13_ssmi, gpm_l1c_f14_ssmi, gpm_l1c_f15_ssmi} \\

SSMI/S &
19.35 GHz, 22.23 GHz, 37 GHz, 91.67 GHz, 150 GHz, 183 $\pm$ 1 GHz, 183 $\pm$ 3 GHz, 187 $\pm$ 6.6 GHz &
2004--current &
\citet{ gpm_l1c_f16_ssmis, gpm_l1c_f17_ssmis, gpm_l1c_f18_ssmis, gpm_l1c_f19_ssmis} \\

AMSU-B &
89 $\pm$ 0.9 GHz, 150 $\pm$ 150.0 GHz, 183.31 $\pm$ 1.0 GHz, 183.31 $\pm$ 7.0 GHz &
2000--2013 &
\citet{gpm_l1c_noaa15_amsub, gpm_l1c_noaa16_amsub, gpm_l1c_noaa17_amsub} \\

MHS &
89 GHz, 157 GHz, 183.31 $\pm$ 1.0 GHz, 183.31 $\pm$ 3.0 GHz, 190.31 GHz &
2007--current &
\citet{gpm_l1c_noaa18_mhs, gpm_l1c_noaa19_mhs, gpm_l1c_metopa_mhs, gpm_l1c_metopb_mhs, gpm_l1c_metopc_mhs}  \\

ATMS &
23.8 GHz, 31.4 GHz, 88.2 GHz, 165.5 GHz, 183.31 $\pm$ 7.0 GHz, 183.31 $\pm$ 4.5 GHz, 183.31 $\pm$ 3.0 GHz, 181.31 $\pm$ 1.8 GHz, 183.31 $\pm$ 1.0 GHz &
2012--2025 &
\citet{gpm_l1c_noaa20_atms, gpm_l1c_npp_atms} \\

GMI &
10.65 GHz, 18.7 GHz, 23.8 GHz, 36.64 GHz, 89 GHz, 166 GHz, 183 $\pm$ 3 GHz, 183 $\pm$ 7 GHz &
2014--2025 &
\citet{gpm_l1c_gpm_gmi, gpm_l1c_gcomw1_amsr2} \\

GOES &
0.64 $\mu$m, 3.9 $\mu$m, 6.9 $\mu$m, 11.2 $\mu$m, 12.3 $\mu$m, 13.3 $\mu$m &
2002--current &
\citet{knapp2018_gridsat_goes, noaa_goes_open_data} \\

Geostationary infrared &
11 $\mu$m &
2000--current &
\citet{noaa_cpc_ir} \\

Goestationary visible and infrared &
0.6 $\mu$m, 6.7 $\mu$m, 11 $\mu$m &
2000--current &
\citet{knapp2011_gridsat} \\
\hline
\end{tabular}
\end{table*}

\subsubsection{Data Preparation}

All experiments are performed on the regular latitude–longitude grid used by the
Prithvi-WxC model with a spatial resolution of 0.625° × 0.5°. Since this grid is
coarser than all other datasets, the IMERG and MRMS precipitation estimates and
the baseline precipitation forecasts are aggregated to coarser grid. IMERG and
MRMS precipitation rates are aggregated to 3-hourly accumulations. Data from
2000 -- 2020 is used to train the Prithvi-Precip model. Data from 2022 is used
for tuning the model design and forecasts from March to August 2025 are used for
the evaluation against the GEOS-FP and AIFS baselines.

%

\section{Results}
\label{sec:results}

\subsection{Design Experiments}

There are multiple design choices to be made for adapting the Prithvi-WxC model
for precipitation forecasts. In order to determine a suitable configuration for
the Prithvi-Precip model, we have performed a number of experiments aimed at
exploring the principal design choices.

\subsubsection{Rollout and Continuous Forecasting}

\begin{figure}[hbpt!]
 \centerline{\includegraphics[width=37pc]{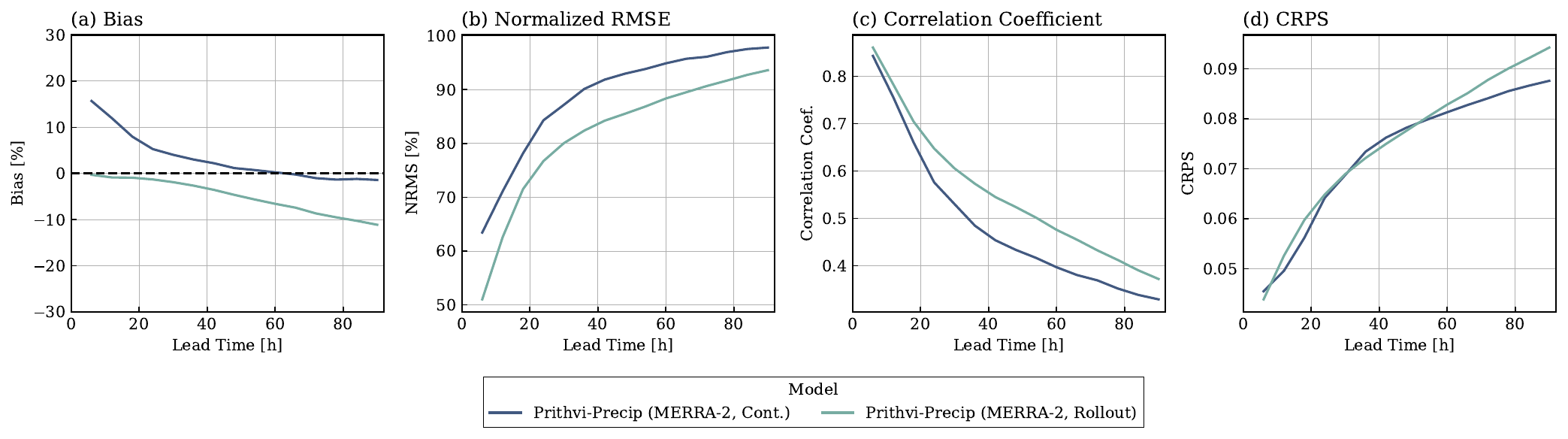}}
 \caption{
   Forecast accuracy for Prithvi-Precip forecasts trained on MERRA-2
precipitation using continuous and autoregressive forecasting. Panel (a) shows
the relative bias, panel (b) the normalized root mean squared error (NRMSE), panel (c)
the correlation coefficient, panel (d) the continuously ranked probability score
(CRPS).
 }\label{fig:forecast_type}
\end{figure}

We first compare autoregressive and continuous precipitation forecasting. For the
autoregressive configuration, the Prithvi-Precip model was trained using six
recursive six-hour forecast steps, corresponding to a maximum lead time of 36
hours. The continuous model was trained to directly predict precipitation at
lead times up to 96 hours. Forecast skill for both approaches is shown in Fig.
\ref{fig:forecast_type}.

In terms of biases, the autoregressive forecast remains nearly bias free up
to a lead time of around 36 hours after which it exhibits an increasing tendency
to underestimate precipitation. Since the model was trained using roll out
up to 36 hours, we suspect that the biases are caused by the model producing
predictions outside the lead times it has encountered during training. The direct
forecast model produces comparably strong positive biases at short lead time that
diminish with increasing lead time.

In terms of all other assessed metrics, the autoregressive model yields higher
forecast accuracy across the full range of lead times. These results suggest
that autoregressive training provides a substantial benefit for precipitation
forecasting, even when the model is constrained only through the predicted
precipitation fields rather than through the evolution of the full atmospheric
state, as is typical for general AI weather prediction systems.

\subsubsection{Precipitation Target Data}
\label{sec:reference_data}
Secondly, we quantify the impact of the target precipitation data used for
training. To this end, we train Prithvi-Precip models using precipitation
estimates from the MERRA-2 reanalysis and IMERG V07 while keeping the model
architecture and training procedure identical. The resulting forecasts are
evaluated against IMERG, MERRA-2 and independent MRMS precipitation estimates
using the temporally independent testing data from 2022.

\begin{figure}[h]
 \centerline{\includegraphics[width=37pc]{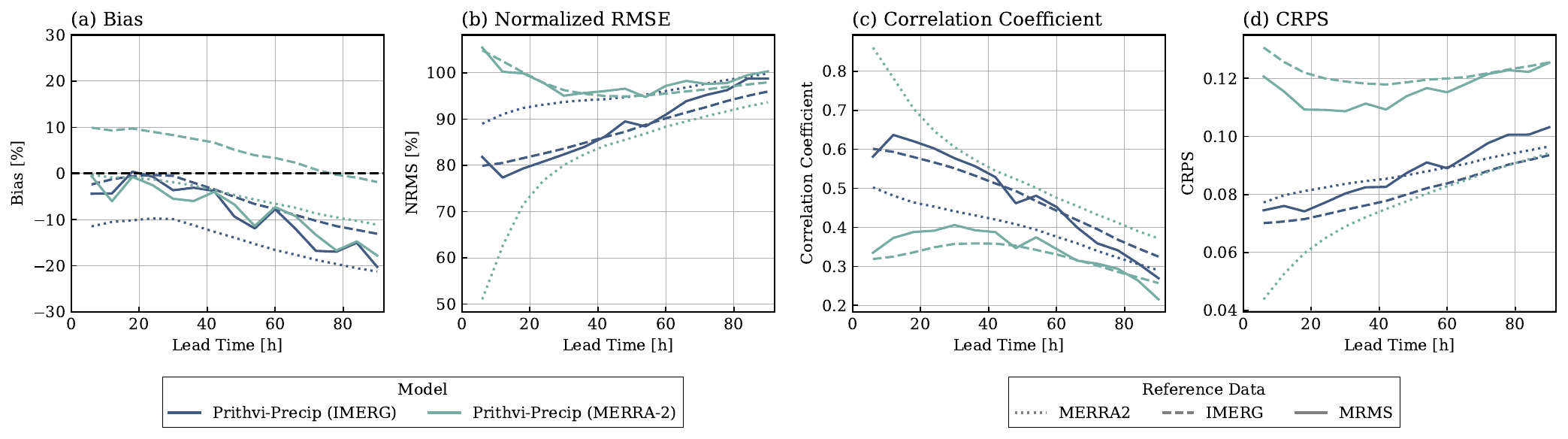}}
 \caption{
   Like Fig.~\ref{fig:forecast_type} but for Prithvi-Precip forecasts trained on MERRA-2 and
   IMERG precipitation target data evaluated against MERRA-2, IMERG, and MRMS precipitation estimates.
 }\label{fig:ref_data}
\end{figure}

The results displayed in Fig.~\ref{fig:ref_data} show that each model achieves
the highest accuracy when it is evaluated against the dataset used during
training but shows degraded performance when evaluated against the independent
MRMS precipitation estimates. This shows the tendency of AIWP forecasts to learn
dataset-specific artifacts that do not generalize to comparison against
independent reference estimates. While testing against temporally independent
data is still often considered sufficient to validate AIWP forecasts, these
results indicate that this may not be sufficient for precipitation estimates.
To simplify the discussion below, we will therefore refer to the temporally
independent data from the same dataset as \textit{in-distribution} data.

Although both Prithvi-Precip models were trained using identical architectures
and optimization procedures, the Prithvi-Precip (MERRA-2) model achieves
substantially higher in-distribution accuracy than the IMERG-based model. This
is likely because MERRA-2 precipitation close to the initialization remains
consistent with and predictable from the MERRA-2 initial conditions, which is
not the case for the IMERG precipitation estimates. However, this advantage
vanishes after about 30 h -- likely as the reanalysis is driven away from its
model trajectory by the assimilated observations. After 30
hours, the in-distribution accuracy of the MERRA-2 and IMERG-based models approach
one another with the in-distribution accuracy of the MERRA-2 forecasts remaining
slightly higher.

When evaluated against the independent MRMS precipitation estimates, the
IMERG-base forecasts consistently outperform the MERRA-2-based forecasts across
all metrics and lead times. This indicates that the IMERG precipitation is more
realistic and therefore leads to better generalization to independent
precipitation estimates.

These results demonstrate that the choice of the target precipitation data has a
substantial impact on forecast behavior and generalization. Although the
MERRA-2-based forecasts yield higher in-distribution skill, the IMERG-trained
forecasts provide substantially better agreement with the independent MRMS
observations. This suggest that the IMERG V07 estimates provide a more
effective training target for precipitation forecasting.

\subsubsection{Model Complexity}

\begin{figure}[h]
 \centerline{\includegraphics[width=37pc]{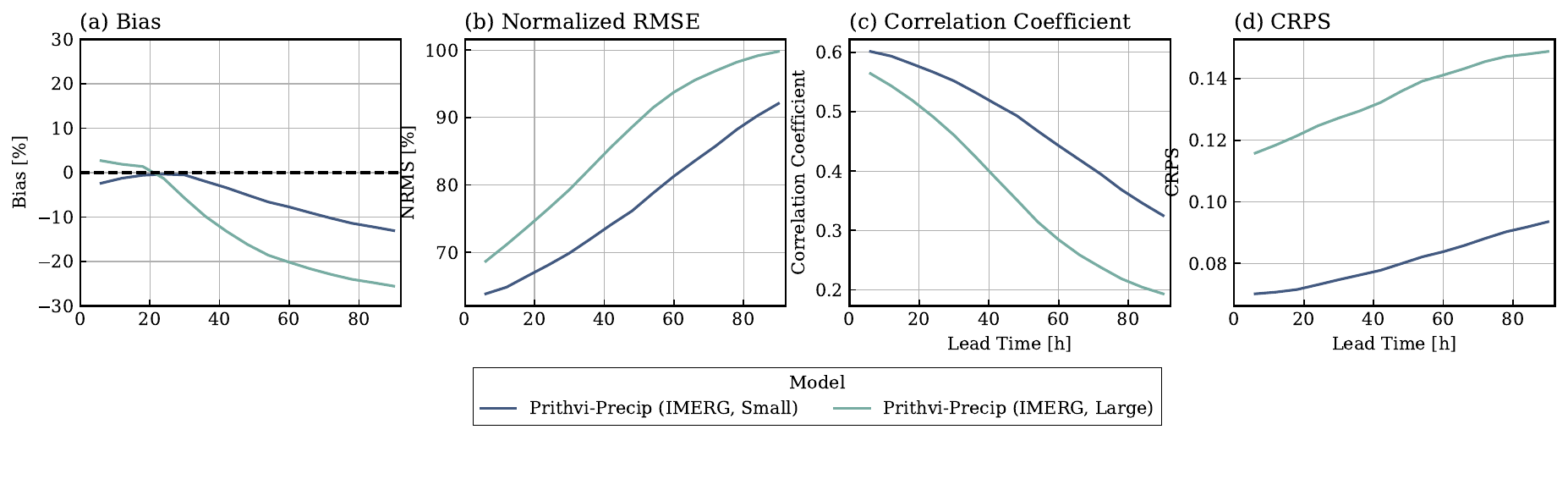}}
 \caption{
   Like Fig.~\ref{fig:forecast_type} but comparing forecast accuracy for the small
   and large Prithvi-WxC model configurations trained using the same number of forecast
   steps.
 }\label{fig:complexity}
\end{figure}

Due to compute constraints all experiments presented up to this point were
performed using a small variant of the Prithvi-WxC model. The original (large)
model described in \citep{schmude_prithvi_2024} contains 2.7 billion parameters,
whereas the small model only contains 280 million parameters. In addition to the
difference in parameter count, the large model was pretrained for autoregressive
forecasting while the small model was not.

To test the impact of the model complexity, we compare Prithvi-Precip models
based on the small and large versions of the Prithvi-WxC model. Due to the
substantially higher computational complexity of the large model, we had to
change the training approach by limiting the rollout to four autoregressive
steps compared to the six steps used for the small model. To make up for the
reduced number of forecast steps, the large model was trained for 150 instead of
100 epochs.

Figure~\ref{fig:complexity} compares the forecast skill of the two models on the
independent testing data. The small model yields consistently improved forecast
accuracy across all assessed lead times. This indicates that the large model
remains undertrained with the given training dataset size and training compute
budget. Since the training already used all available IMERG training data prior
to 2020, we cannot increase the training dataset size. Furthermore, since we found the small
model overfit when trained for more than 100 epochs, we consider it unlikely
that its accuracy is limited by model complexity. We therefore adopt the small
model for the Prithvi-Precip forecasts.

\subsection{Observation-Driven Precipitation Forecasts}

\begin{figure}[h]
 \centerline{\includegraphics[width=37pc]{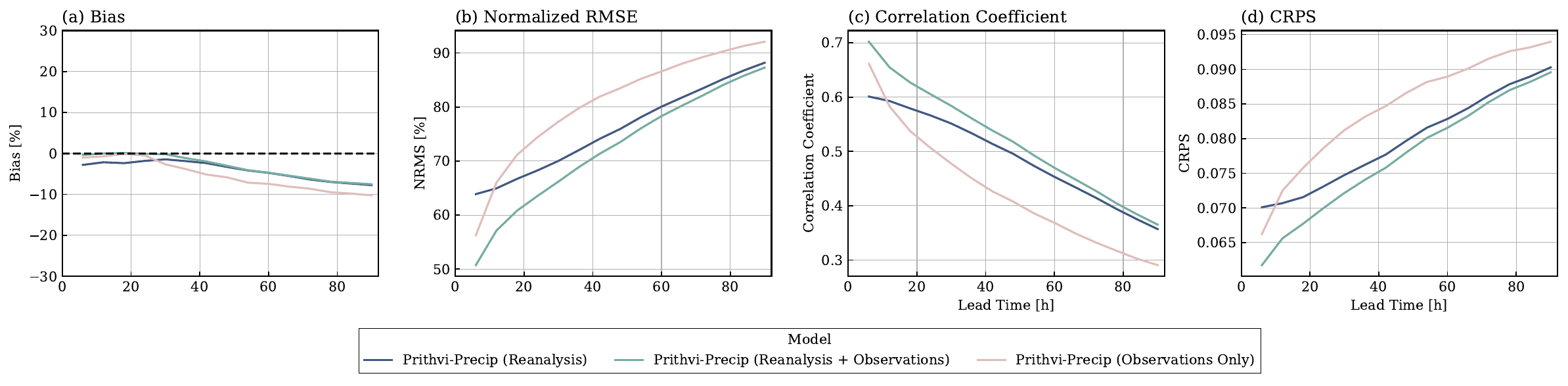}}
 \caption{
   Like Fig.~\ref{fig:forecast_type} but comparing Prithvi-Precip forecasts trained on IMERG
   precipitation and ingesting raw satellite observations. All forecasts are derived from the
   Prithvi-Precip model with all inputs and dropping either the analysis and observation inputs.
 }\label{fig:obs_forecasts}
\end{figure}

Based on the experiments presented above, we use the Prithvi-Precip based on the
small Prithvi-WxC model with autoregressive forecasting for the
observation-driven forecasts. Figure~\ref{fig:obs_forecasts} compares forecast
skill for three forecast configurations: using only reanalysis inputs, using
only satellite observations, and using the combined reanalysis and satellite
inputs. Incorporating the satellite observations into the forecast leads to
robust improvements in forecast skill across normalized root mean square error
(NRMS), correlation, and continuously ranked probability score (CRPS), with the
largest gains occurring at short lead times up to approximately 40 hours after
which the benefits diminish. The observation-only forecasts achieve high
accuracy at 6-hour lead time but decrease rapidly in accuracy falling below that
of the reanalysis-only forecasts already after 12 hours. After that, the
accuracy of the observation-only forecasts remains substantially lower than that
of the other forecasts.

\begin{figure}[h]
 \centerline{\includegraphics[width=37pc]{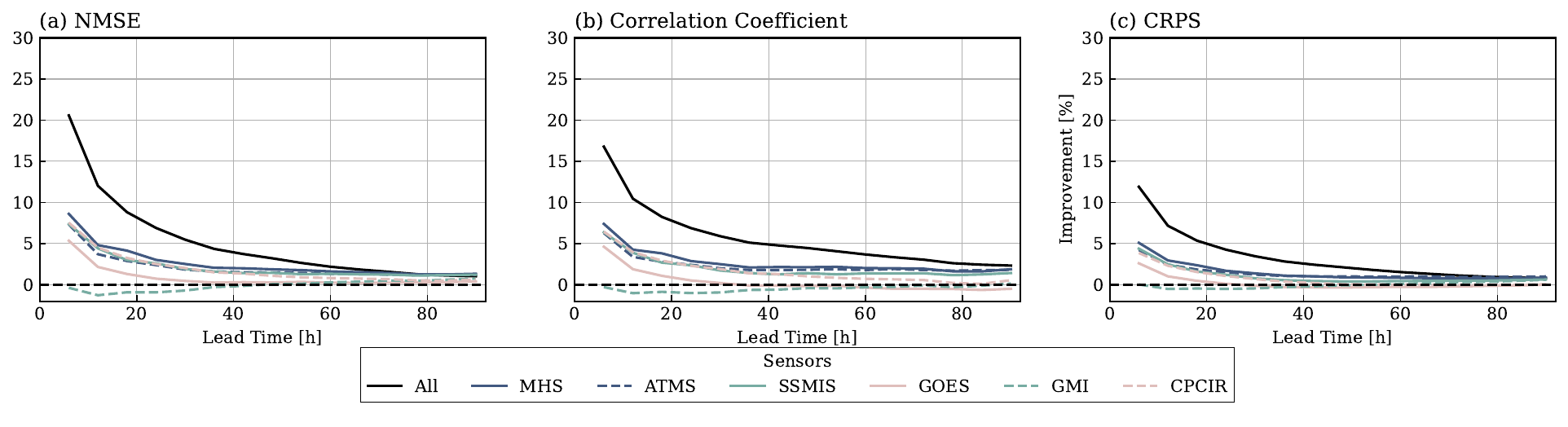}}
 \caption{
  Assessment of sensor impact on forecast skill. Panels (a)–(d) show the relative
  improvement over the reanalysis-only forecast in normalized root mean squared error
  (NRMSE), linear correlation coefficient, and CRPS, respectively, when the model
  is run using all available observations or only observations from a specific
  sensor type.
 }\label{fig:sensor_impact}
\end{figure}

To further investigate the characteristics of the observation-driven
precipitation forecasts, we evaluate the impact of each sensor type on the
forecast accuracy. To this end we run the forecasts using only observations from
each single sensor type. Figure~\ref{fig:sensor_impact} shows the relative
improvements in NRMSE, correlation coefficient, and CRPS for forecasts driven
with observations from different sensors compared to the forecast run using only
reanalysis inputs. Except for GMI, all sensors have positive impact on the
forecast accuracy. PMW observations from operational meteorological sensors
(MHS, ATMS and SSMI/S) and single-channel IR observations all have comparable
positive observation impact. Multi-spectral geostationary observations from GOES
have a lower impact -- likely because they are only available over the Americas
and adjacent parts of the Atlantic and Pacific.

Observations from GMI are found to have weakly negative impact that transitions
to weakly positive for lead times exceeding 60 hours. We suspect that this may
be an artifact of the experiment setup, as the model was not explicitly trained
to use  GMI observations alone. We have performed an observation denial experiment
that did not yield improved forecasts when GMI observations were withheld.
This indicates that GMI observations may be redundant but do not negatively affect
the forecast accuracy when combined with observations from other sensors.

\begin{figure}[h]
 \centerline{\includegraphics[width=37pc]{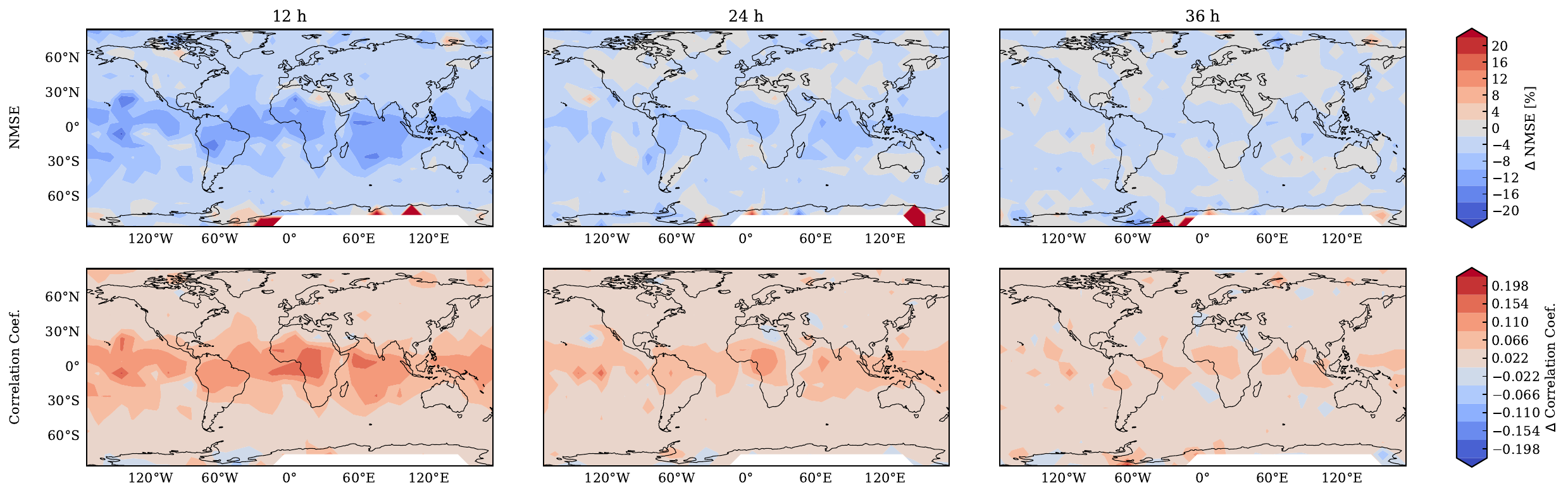}}
 \caption{
   Geographical distribution of the impact of satellite observations on the
normalized root mean squared error (NRMS) and correlation coefficient. The first
row shows the impact on the NRMSE at 12, 36, and 48 hours lead time. The second
row shows the corresponding results for the correlation coefficient.
 }\label{fig:geographical_impact}
\end{figure}

Finally, we examine the geographical distribution of the forecast improvements
resulting from the direct integration of satellite observations.
Figure~\ref{fig:geographical_impact} displays the spatial distribution of the
average relative improvement in NRMSE and linear correlation at three forecast
lead times. The impact exhibits a pronounced geographical structure,
particularly at shorter lead times, with the largest improvements occurring
across the tropics and subtropics. We hypothesize that this pattern reflects the
prevalence of convective precipitation processes in these regions, which are
often inadequately represented in the initial conditions provided by the MERRA-2
reanalysis. By directly incorporating information from raw satellite
observations, Prithvi-Precip is able to compensate for limitations of
conventional data assimilation systems and provide a more accurate
representation of precipitation-producing weather systems.

\subsection{Comparison against Operational Models}

To assess the operational relevance of the Prithvi-Precip model, we compare it
against operational precipitation forecasts from the GOES-FP and AIFS forecast
systems. The Prithvi-Precip forecasts are run in three configurations: using
only analysis data, using only satellite observations, and using analysis and
observation data. We run forecasts at 0 Z and 12 Z from March 2025 up to and
including August 2025. The limitation to 0 Z and 12 Z initialization times was
imposed by the availability of the GOES-FP forecasts. The discrete
initialization times interact with the diurnal cycle of precipitation causing
notable oscillations in the forecast accuracy that become particularly notable
in the evaluation against the MRMS and INMET gauge data. The GOES-FP and AIFS
forecasts were spatially aggregated to the MERRA-2 grid for the comparison.

We compare the forecasts using IMERG V07, MRMS, and INMET precipitation
estimates as reference data. The evaluation against IMERG V07 will naturally
favor the Prithvi-Precip model since it was trained the data. However, these
results are include here as a reference to separate model capabilities from
the impact of the training data accuracy.

\subsubsection{IMERG V07}

\begin{figure}[h]
 \centerline{\includegraphics[width=37pc]{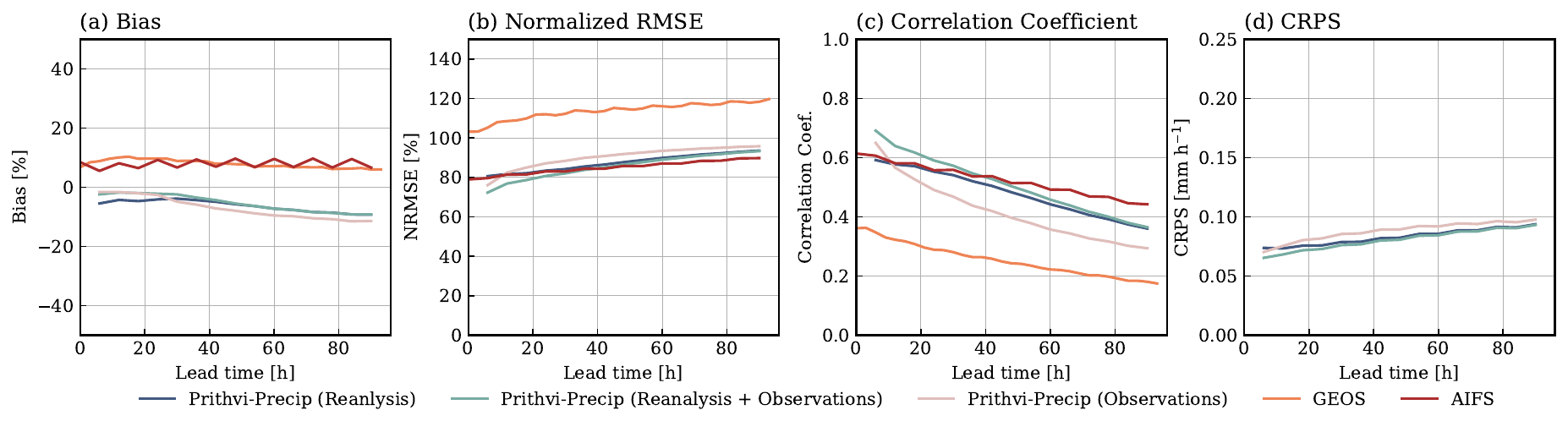}}
 \caption{
   Evaluation of Prithvi-Precip, GEOS-FP, and AIFS forecasts against global precipitation
   estimates from IMERG V07.
 }\label{fig:evaluation_imerg}
\end{figure}

The forecast accuracy for the  GEOS FP, AIFS, and three Prithvi-Precip
configurations are shown in Fig.~\ref{fig:evaluation_imerg}. When assessed using
IMERG V07 precipitation estimates, the Prithvi-Precip forecasts driven using
both reanalysis data and raw satellite observations yield the highest accuracy
up to lead times of 40 hours. After that, they fall below the accuracy of the
AIFS forecasts. The reanalysis-only forecasts are less accurate and remain at
the level of the AIFS forecasts but fall below them after lead times exceeding
24 h. The observation-only forecasts are more accurate than the AIFS baseline
for the 6-hour forecast but fall below it already after 12 hours. All
Prithvi-Precip forecasts substantially improve upon the GOES-FP baseline
forecasts across the full forecast window.

\subsubsection{MRMS}

\begin{figure}[h]
 \centerline{\includegraphics[width=37pc]{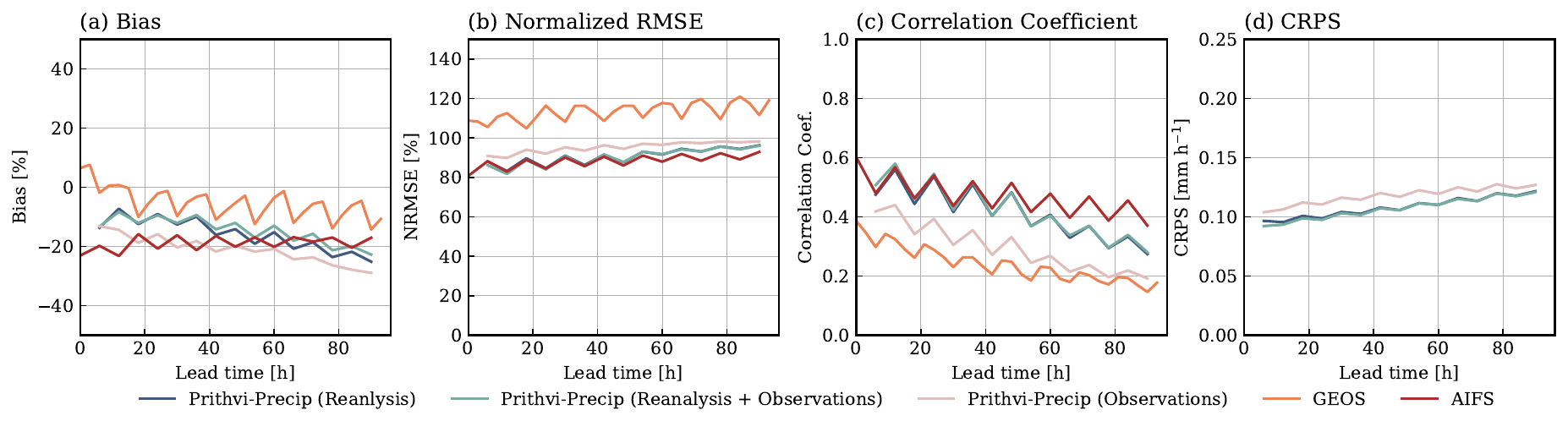}}
 \caption{
   As Fig.~\ref{fig:evaluation_imerg} but with the forecasts evaluated against
MRMS gauge-corrected precipitation estimates from ground-based radars over
CONUS.
 }\label{fig:evaluation_mrms}
\end{figure}

Figure~\ref{fig:evaluation_mrms} shows the forecast accuracy evaluated against
ground-based radar measurements over CONUS. As expected, the relative accuracy
of the Prithvi-Precip forecasts is lower than for the evaluation against the
IMERG data. The reanalysis-and-observation-driven Prithvi-Precip forecast
retains a minimal advantage over the AIFS forecast up to lead times of 18 hours
and remains at the level of the AIFS forecast up to about 36 hours. The
reanalysis-only Prithvi-Precip forecast achieves accuracy similar to AIFS up to
36 h after which they closely track the accuracy of the
reanalysis-and-observation forecasts and become less accurate than the AIFS
forecast.

It is notable that ingesting satellite observations yields only very small
improvements in forecast accuracy compared to the reanalysis-only
forecast. We suspect that the reason for this is the broad availability of
meteorological observations over CONUS that are assimilated into the MERRA-2
reanalysis used to initialize the forecasts. Since this increases the accuracy
of the initial state, there is little additional information that the model
could extract from the satellite observations to improve the precipitation
forecast.

\subsubsection{INMET Gauges}

\begin{figure}[h]
 \centerline{\includegraphics[width=37pc]{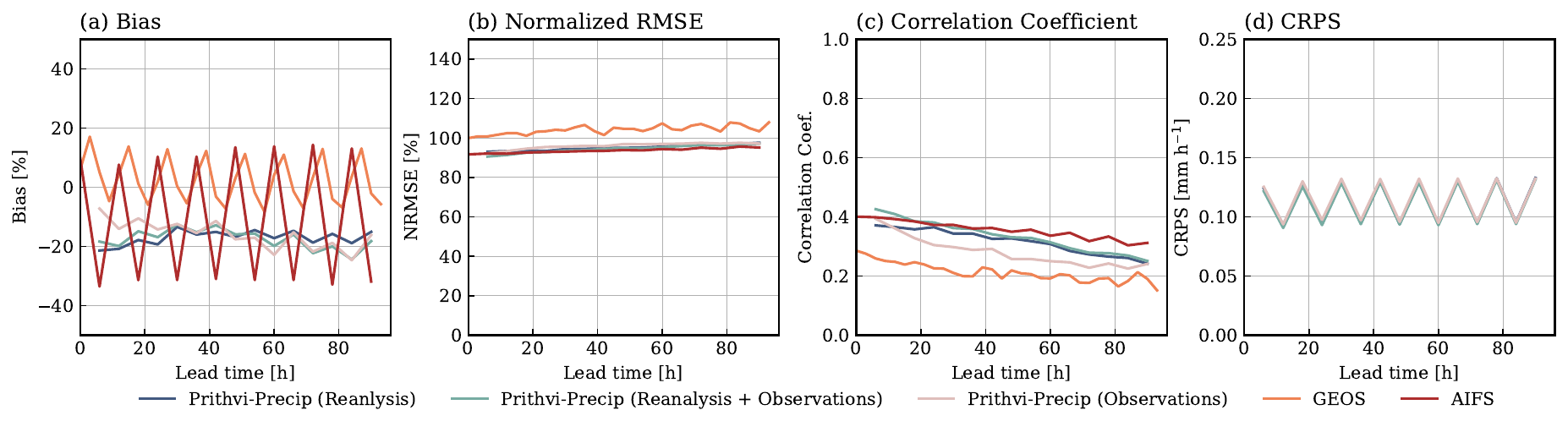}}
 \caption{
   As Fig.~\ref{fig:evaluation_imerg} but with the forecasts evaluated against
hourly gauge measurements from the Instituto Nacional de Meteorologia of Brazil.
 }\label{fig:evaluation_inmet}
\end{figure}

Finally, we compare the forecasts to gauge measurements over Brazil to provide
an additional assessment against independent precipitation estimates over
tropical and sub-tropical land. The results -- shown in
Fig.~\ref{fig:evaluation_inmet} -- indicate  lower overall accuracy than over
the mid-latitudes. This is likely due to the reference data being derived directly
from gauges that are less representative at the MERRA-2 grid scale due to their
point-measurement character. Additionally, it  reflects a reduced predictability
of precipitation at low latitudes at the targeted lead times \citep{keane_2025}

Over Brazil, the impact of ingesting satellite observations in addition to the
reanalysis initial conditions yields a larger relative impact than over CONUS.
The reanalysis-and-observation-based forecasts achieve slightly higher forecast
accuracy than the AIFS at 6 and 12 hour lead times and again fall below the
accuracy of the AIFS at lead times exceeding 40 hours. The reanalysis-only forecasts
remain below the accuracy of the AIFS forecasts over Brazil likely indicating
less accurate initial conditions compared to CONUS.

\section{Summary and Conclusions}
\label{sec:summary_and_conclusions}

This study introduced the Prithvi-Precip AI precipitation forecast model.
Prithvi-Precip was developed by finetuning the Prithvi-WxC foundation model for
medium-range precipitation forecasting. To identify effective strategies for
improving AI-based precipitation forecasts, we assessed key design choices
related to the training methodology, precipitation reference data, model
complexity, and the integration of satellite observations.

Our experiments demonstrate that autoregressive finetuning substantially
improves forecast accuracy compared to continuous forecasting, even when the
model is trained only on the evolving precipitation field and not the full
atmospheric state. Furthermore, forecast accuracy generally exhibited enhanced
degradation at lead times exceeding the training rollout length. These findings
suggest that AI foundation models for weather forecasting systems should be
optimized for long autoregressive rollouts and highlight the importance of
memory- and compute-efficient architectures that enable such training.

A second key finding concerns the choice of precipitation targets. While models
trained using MERRA-2 precipitation achieve higher agreement with temporally
independent MERRA-2 estimates, they exhibit substantially reduced skill when
evaluated against independent precipitation observations from ground-based
radar. In contrast, training on IMERG V07 precipitation estimates yields
significantly better generalization to independent observations. These results
demonstrate that the accuracy of the precipitation targets used during training
has a direct impact on the quality of the resulting forecasts.

We also investigated the impact of model complexity by comparing a 280 million
parameter model with a 2.7 billion parameter variant. Despite its substantially
larger capacity, the larger model did not improve forecast skill under the
available training budget. We suspect that this is due to the compute
limitations that forced us to reduce the number of rollout steps as well as the
fixed training length in the experiment. The larger model was likely remained
undertrained and could likely be brought to same or higher forecast accuracy
with longer training. However, given that we achieved competitive results with
the smaller model as well as the computational constraints common in geophysical
research, these findings highlight the value of foundation models that are
available in a range of model sizes.

The central methodological innovation of this work is a sensor-agnostic
framework for incorporating satellite observations directly into AI weather
prediction models. The approach represents each observed channel as an
independent observation layer and augments it with metadata describing the
sensor characteristics and sampling geometry. By treating sensor properties as
input features rather than relying on sensor-specific model architectures, the
framework can continuously learn from observations acquired by different sensors
and sensor generations while remaining scalable to a large and heterogeneous
observing system.

To increase robustness, the model is trained with random dropout of both
observation and model inputs. This enables the forecasting system to operate
with complete, partial, or entirely missing observations and allows forecasts to
be generated directly from observations without requiring a conventional data
assimilation system. Such a capability could substantially simplify the
operational deployment of AI forecasting systems.

Combining two principled interventions—training on higher-quality precipitation
estimates and directly ingesting satellite observations—substantially reduced
the performance gap between the GEOS-FP baseline and AIFS precipitation
forecasts. While Prithvi-Precip achieves higher agreement with IMERG
precipitation estimates than AIFS at lead times up to approximately 40 hours,
these improvements translate only partially to evaluations against independent
observations over CONUS and Brazil, where gains are primarily observed at 6- and
12-hour lead times. At longer lead times the AIFS still remains substantially more
accurate, which is potentially due to the differences in the underlying data
assimilation systems. The IFS employs one of the world most advanced data
assimilation system whereas MERRA-2 is based on a much simpler 3-DVar system.

Despite these remaining limitations, the results identify two promising pathways
for further improving AI precipitation forecasting. The first is improving the
quality and quantity of precipitation training data. Although IMERG V07 provides
a clear improvement over MERRA-2 precipitation estimates, it remains affected by
retrieval uncertainties. The upcoming IMERG V08 product is expected to provide
substantially more accurate precipitation estimates through the adoption of
machine-learning-based retrieval algorithms \citep{pfreundschuh_gprof-nn_2022,
pfreundschuh_gprofir_2026}.

The second pathway is the direct use of satellite observations. Our results
demonstrate that satellite observations provide information beyond that
contained in conventional analyses and can help improve forecast accuracy,
particularly at short lead times and in the tropics and subtropics. Because the
current implementation ingests only a subset of available satellite
observations, substantial opportunities remain for further improvement.

Taken together, these results suggest that future advances in AI precipitation
forecasting may be achieved by improving the quality of precipitation training
targets and more fully exploiting the global satellite observing system. Both
directions offer clear opportunities for improving forecast skill and may enable
AI-based forecasting systems to further improve upon conventional NWP
approaches.

\clearpage
\acknowledgments

The work of Simon Pfreundschuh work has been supported by NASA grant 80NSSC22K0604.

\datastatement

The implementation of the Prithvi-Precip model and the code to recreate training
and testing data from publicly available data sources are available from a
public repository \citep{prithvi_precip}.








%




\bibliographystyle{ametsocV6}
\bibliography{references}

\end{document}